\documentclass[fleqn,10pt]{wlscirep}
\usepackage[utf8]{inputenc}
\usepackage[T1]{fontenc}
\title{Social media emotion macroscopes reflect emotional experiences in society at large}

\author[1,2,3*]{David Garcia}
\author[1,2,3]{Max Pellert}
\author[1,2]{Jana Lasser}
\author[1,2,3,4]{Hannah Metzler}

\affil[1]{Graz University of Technology, Institute of Interactive Systems and Data Science, Department of Computer Science and Biomedical Engineering, 8010 Graz, Austria}
\affil[2]{Complexity Science Hub Vienna, 1080 Vienna, Austria}
\affil[3]{Medical University of Vienna, Center for Medical Statistics, Informatics and Intelligent Systems, 1090 Vienna, Austria}
\affil[4]{Medical University of Vienna, Department of Social and Preventive Medicine, Center for Public Health, 1090 Vienna, Austria}
\affil[*]{dgarcia@tugraz.at}

\begin{abstract}

Social media generate data on human behaviour at large scales and over long periods of time, posing a complementary approach to traditional methods in the social sciences\cite{Salganik2019}.
Millions of texts from social media can be processed with computational methods to study emotions \cite{Mohammad2018,Goldenberg2020} over time and across regions \cite{Golder2011,Dodds2011}.
However, recent research has shown weak correlations between social media emotions and affect questionnaires at the individual level \cite{Beasley2016,Elayan2020} and between static regional aggregates of social media emotion and subjective well-being at the population level \cite{Jaidka2020}, questioning the validity of social media data to study emotions.
Yet, to date, no research has tested the validity of social media emotion macroscopes to track the temporal evolution of emotions at the level of a whole society.
Here we present a pre-registered prediction study that shows how gender-rescaled time series of Twitter emotional expression at the national level substantially correlate with aggregates of self-reported emotions in a weekly representative survey in the United Kingdom.
A follow-up exploratory analysis shows a high prevalence of third-person references in emotionally-charged tweets, indicating that social media data provide a way of social sensing the emotions of others \cite{Galesic2021} rather than just the emotional experiences of users.
These results show that, despite the issues that social media have in terms of representativeness \cite{Lazer2021} and algorithmic confounding \cite{Wagner2021}, 
the combination of advanced text analysis methods with user demographic information in social media emotion macroscopes can provide measures that are informative of the general population beyond social media users.

\end{abstract}

\begin{document}

\maketitle
%
%
\thispagestyle{empty}


In our digital society, new technologies for social interaction are playing a key role in capturing and shaping our emotional experiences. Affect plays an central role in several important phenomena \cite{Dukes2021rise}, including affective polarization across political identities \cite{Iyengar2019,Finkel2020}, emotional components in misinformation \cite{Martel2020}, and sentiment in online attention and engagement \cite{Rathje2021}. At the same time, social technologies generate data on social interaction that has the potential to capture emotional expression at new scales and resolutions. This has motivated applications of social media data to affective research questions such as mental health \cite{Chancellor2020}, emotional well-being \cite{Jaidka2020}, anxiety \cite{Elayan2020}, collective emotions \cite{Garcia2019}, and emotion regulation \cite{Fan2019}. An especially powerful approach is the implementation of emotion macroscopes that can aggregate affective information at large scales and fast temporal resolutions, often relying on text analysis \cite{Golder2011} or mental health tracker data \cite{Yarrington2021}. However, social media data is not designed for behaviour research, bringing concerns about algorithmic and preformative behaviour issues \cite{Salganik2019,Wagner2021}, as well as sampling biases that make social media users not representative of the population at large in terms of demographics and other relevant attributes \cite{Lazer2021}.

Testing the validity of social media emotion macroscopes as a measurement of aggregates of emotional experiences has remained an elusive task due to three challenges: scarcity of frequent representative surveys of emotions, the technical barriers to analyze large-scale longitudinal geo-located data from social media, and the low precision of emotion detection methods from social media text. Here we address these three challenges in a pre-registered validation study of social media macroscopes of several emotional states. We use a comprehensive dataset of social media text from the United Kingdom and apply both established and advanced text analysis methods to measure emotion aggregates. Results are then compared with two years of emotion data from a representative survey in the UK.

We present a pre-registered study testing that Twitter aggregated emotion timelines positively correlate with weekly emotion reports when applying dictionary-based methods and taking gender into account (more details about pre-registration in the Supplementary Information).
The first part of our pre-registration describes a test of our hypotheses for historical data up to October 2020. The second part of the pre-registration postulates the same hypotheses and analysis as a prediction for a future time period. This way we test the generalizability of our results across time, making a proper prediction rather than doing retrodictive data analysis alone \cite{Gayo2013}.
Our analysis covers the period between June 2019 and June 2021, including a total of more than 1.5 Billion tweets posted by users in the United Kingdom (see Materials and Methods for more information).
We designed the study to test two approaches to emotion detection from text: the most popular dictionary-based method with emotion word lists in English \cite{Pennebaker2015}, and a state-of-the-art supervised classifier that we trained against a large corpus of tweets with annotated emotions \cite{Mohammad2018} (see Methods).

The fraction of tweets posted by men is above 60\%, in line with the higher visibility of men on Twitter \cite{Nilizadeh2016}. Because of this, gender-agnostic measures have the risk of overweighting male voices on Twitter. This can be a source of systematic error given the known gender differences in emotional experiences \cite{chaplin_gender_2013, mclean_brave_2009, whittle_sex_2011}. We use gender information about the users in our sample to calculate gender-rescaled emotional expression indices that are more representative of the population of the UK.


The measurements of sadness in the survey and on Twitter are consistently correlated between historical and predicted periods of our pre-registration, as shown in Figure 1. The dictionary approach achieves substantial positive correlations (0.69 in the historical period and 0.672 in the prediction period) as well the supervised method (0.636 in the historical period and 0.653 in the prediction period).
Similarly, for the case of \emph{scared} in the survey versus anxiety or fear in Twitter, illustrated in Figure 1, shows strong correlations in the historical period (0.78 for dictionary method and 0.793 for supervised method). As preregistered, the dictionary method is positively correlated in the prediction period with a coefficient of 0.471, but the supervised method has a weaker correlation close to 0.30. A summary and statistical details of these correlations are reported on Table 1.

\begin{figure}[ht]
\centering
\includegraphics[width=0.7\linewidth]{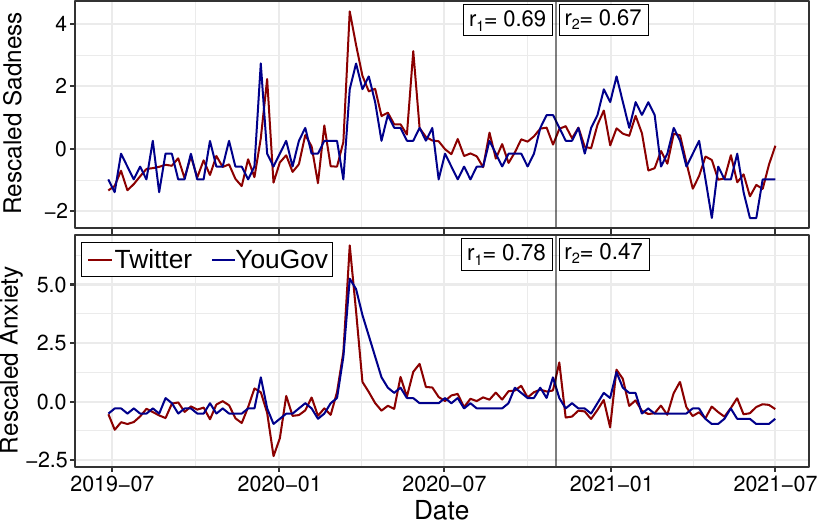}
\caption{Top: Time series of weekly proportion of \textit{sad} responses in YouGov and gender-rescaled sadness score on Twitter based on dictionary analysis. Bottom: Time series of weekly proportion of \textit{scared} responses in YouGov and gender-rescaled anxiety score on Twitter based on dictionary analysis. The grey vertical line marks the start of the prediction period. Reported correlation coefficients between YouGov and Twitter time series are calculated for the historical period ($r_1$) and for the prediction period ($r_2$), more details are reported on Table \ref{tab:1}.
}

\label{fig:1}
\end{figure}

For the case of happiness in the survey, Fig 2 shows the comparison with the gender-rescaled values for joy in the supervised method. This method achieves correlations above 0.5 in both the historical and prediction periods.
On the other hand, the dictionary method for positive affect barely correlates during the historical period and has a correlation close to zero in the prediction period, contradicting our pre-registered hypothesis and indicating that emotion macroscopes of positive emotions work better when using supervised methods.
These results are robust when considering autocorrelation and heteroskedasticity for all cases except for the last case of dictionary-based positive emotions. The case of dictionary-based anxiety also shows some signs of non-stationarity due to the large spike in the time series at the onset of the pandemic, as already reported with similar methods \cite{Metzler2021}, but results are still comparable when using robust estimators that consider outliers and in permutation tests (see SI for more details).
Signals measured with the supervised method are comparable to the dictionary-based method for sadness and anxiety, but for the case of the emotion \emph{happy} in YouGov, the supervised method gives significantly stronger positive correlations than the dictionary-based method in both the historical and prediction periods. This can be attributed to the lack of specificity of positive affect as a class in emotion dictionaries, akin to general metrics of valence or happiness.

\begin{figure}[ht]
\centering
\includegraphics[width=0.7\linewidth]{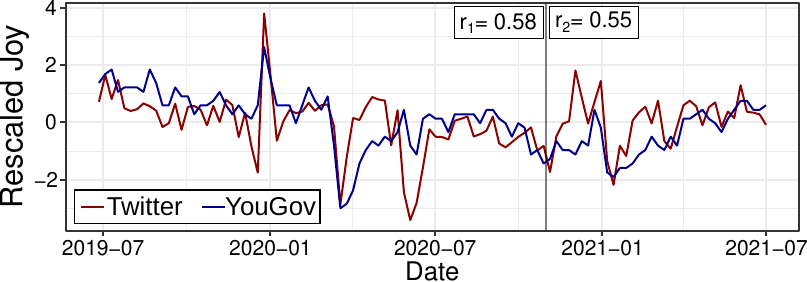}
\caption{Time series of weekly proportion of \textit{happy} responses in YouGov and gender-rescaled joy score on Twitter using our supervised classifier. The grey vertical line in panels marks the start of the prediction period. Reported correlation coefficients between YouGov and Twitter time series are calculated for the historical period ($r_1$) and for the prediction period ($r_2$).
}
\label{fig:1.2}
\end{figure}

\begin{table}[ht]
\centering
\begin{tabular}{|c|c|c|c|c|c|}
\hline
Emotion (survey) & Twitter signal & $r_1$ (historical) &  $r_2$ (predicted)    \\
\hline
Sad & LIWC sad & $0.688^{***} \quad [0.542, 0.794]$ & $0.672^{***} \quad [0.437, 0.821]$\\
Sad & RoBERTa sad & $0.636^{***} \quad [0.472, 0.757]$ & $0.653^{***} \quad [0.408, 0.81]$\\
Scared & LIWC anxiety & $0.78^{***} \quad [0.668, 0.857]$ & $0.471^{**} \quad [0.163, 0.695]$\\
Scared & RoBERTa fear & $0.793^{***} \quad [0.687, 0.866]$ & $0.295^{\cdot} \quad [-0.042, 0.572]$\\
Happy & LIWC positive & $0.298^{*} \quad [0.069, 0.497]$ & $0.043 (n.s.) \quad [-0.295, 0.371]$\\
Happy & RoBERTa joy & $0.576^{***} \quad [0.396, 0.713]$ & $0.551^{***} \quad [0.267, 0.747]$\\
\hline
\end{tabular}
\caption{\label{tab:1} Correlation coefficients of the weekly percentage of emotions in YouGov and a LIWC or RoBERTa estimator based on gender-rescaled Twitter data. 95\% Confidence Intervals are reported. $^{\cdot} p<0.1, ^{*} p<0.05, ^{**} p<0.01, ^{***} p<0.001$}
\end{table}


We further studied the total of 12 emotional states tracked by the YouGov survey, describing an exploratory analysis plan in our pre-registration to study the predictive power of future tweets that contain explicit reports of that emotion, such as "I am sad" for sadness. Figure 3 shows the correlation coefficient for each emotional state in the gender-rescaled case, revealing a tendency towards positive correlations. Four negative states (\emph{scared}, \emph{frustrated}, \emph{stressed} and \emph{bored}) show correlations comparable to the results using supervised classifiers for the cases of scared and sadness, with significant positive correlation coefficients between 0.5 and 0.7. Weak and non-significant correlations happen for emotional states that are not shared frequently on Twitter, such as content or apathetic. All these correlations are comparable in estimates without gender rescaling (see SI for more details).

\begin{figure}[h]
\centering
\includegraphics[width=0.5\linewidth]{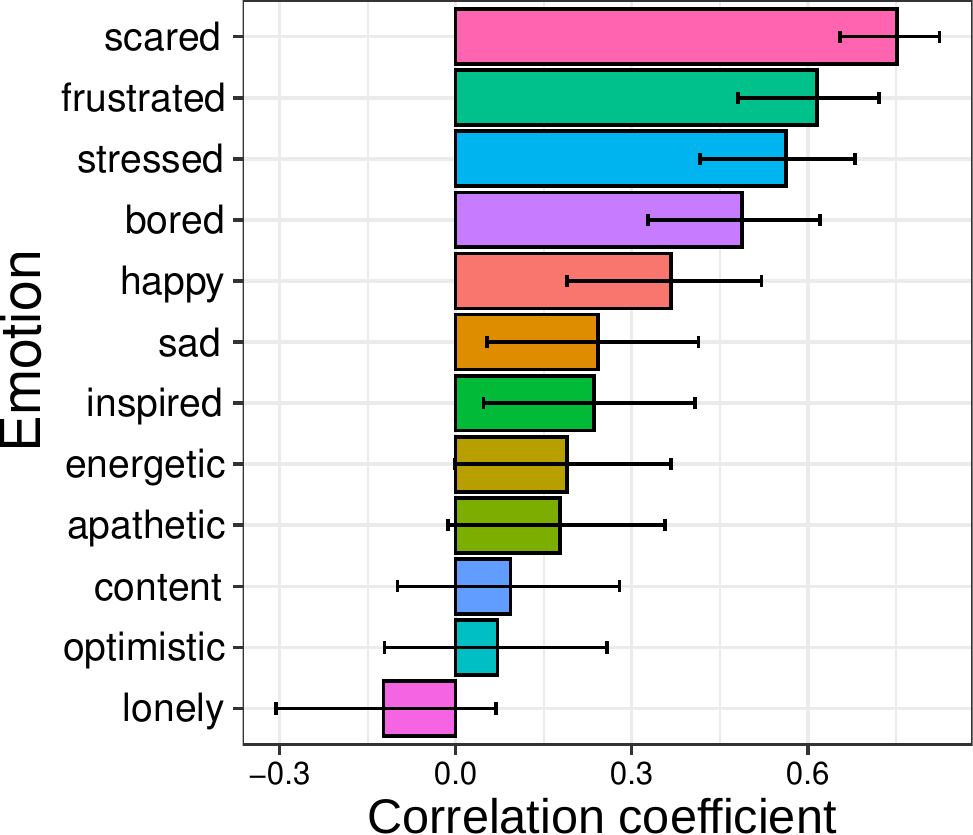}
\caption{Correlation coefficients between the percent of weekly responses for an emotion in the YouGov survey and the weekly volume of tweets explicitly reporting that emotion. Error bars show 95\% confidence intervals.}
\label{fig:2}
\end{figure}

In our analyses, we have found correlations between survey data and emotion macroscopes with values approximately between 0.5 and 0.8, especially when applying supervised emotion detection. 
These correlations are comparable with the correlations between parallel weekly surveys of subjective information.
For example, for the case of US pre-election polls, parallel weekly polls had a correlation of 0.66 for the voting intention for Joe Biden and 0.53 for the voting intention for Donald Trump (see SI for more details). This illustrates that the correlations we found here are in the range of the correlations between surveys, positioning social media as far from perfect but suitable as a complementary source of information on emotion.
Our work is the first to measure the error made when using social media as a macroscope of emotions. Future research can include these error measurements in statistical analyses to test if conclusions based on social media data are robust to measurement error.

Given the different composition of Twitter users compared to the general population in terms of demographics and ideology \cite{Mellon2017}, the positive and consistent correlations reported above are somewhat surprising. There are two possible explanations for these positive correlations. One is the translatability of affective research \cite{Salganik2019}, especially in contrast to other cases like public opinion, which has been shown challenging to track with social media sentiment \cite{Conrad2019}.
The biological basis of emotions can be stronger than the demographic and ideological factors that decide whether a person is active on Twitter.
A second explanation is that, when detecting emotions online, we might not be necessarily measuring the emotional state of the users that posted the tweets but the emotions of others around them and the stimuli that triggered them. This way, social media users could serve as a way of social sensing a larger population that could more closely match the composition of society at large. Despite social sensing being subject to sampling biases due to homophily \cite{Galesic2021}, in the case of social media it might be increasing the representativeness of the sample we study. We tested this idea by comparing third person references in emotional versus non-emotional tweets. We found that the likelihood of a tweet containing a third-person reference is 74.9\% larger for tweets that contain sadness terms than for tweets that do not, 62.1\% larger for tweets that contain fear terms than for tweets that do not, and 35\% larger for tweets that contain positive affect terms than for tweets that do not. A similar analysis with named entity recognition for mentioned persons gives similar results (see SI for more details). While social media emotions are unlikely to be only social sensing, this third-person focus suggests a significant component of mentioning other individuals when talking about emotions on Twitter.


Our results show that social media macroscopes of various emotions can correlate with survey results when including broad location filters and considering gender of social media users. An advanced supervised method performs better than dictionary methods, but the correlations with dictionary methods in the case of negative emotions are still substantial despite the simplicity of that approach. However, these results do not mean that all social media macroscopes of emotions will always work in all populations. Research using these methods should continue developing reliable emotion detection, location filtering, and gender considerations before building macroscopes. These then need to be validated against other social science methods, such as surveys or other kinds of self-reports.
Second, social media drift \cite{Salganik2019} is an issue that can always limit the applicability of results like ours. Even though we used two years of data and performed a prediction study to test the temporal robustness of our results, drift can happen at longer timescales and change how society is represented on social media. Research needs to include adaptive approaches, such as updating statistical models, to avoid falling into the typical pitfalls of computational social science in terms of temporal validity \cite{Munger2019,Lazer2021}.
Beyond these limitations, our results provide the insight that social sensing is a plausible explanation for the correlations we observed.

\section*{Materials and Methods}

We accessed the full historical record of public Twitter data with Brandwatch (previously known as Crimson Hexagon). We analyzed original tweets (excluding retweets) posted by users with a location detected to be in the United Kingdom by the location detector of Brandwatch, which uses user profile information and geo-tagged data to estimate the country in which a user resides \cite{Brandwatch2020}. We also excluded users with less than 100 followers to remove spammers and bots and users with more than 100,000 followers to remove mass media and large organizations. This way, the final sample we analyze contains $1,535,210,474$ tweets posted in the two-year period between July 2019 and June 2021. Our sample is based on tweets that were still available in July 2021, ignoring tweets that had been removed due to accounts being suspended or users deleting Tweets. This way we leverage Twitter's bot detection methods to ignore users who were eventually suspended by Twitter after posting tweets. In addition, this approach respects user decisions to remove their tweets from public display. User gender is based on Brandwatch's classifier, which is  determined from self-reported information on user profiles \cite{Brandwatch2020b}. Not all users' gender can be reliably classified this way, with gender being detected for 55\% of tweets in our sample. However, does not apply this way institutional or non-human accounts, and thus this filtering discards some tweets that could introduce noise in the measurement. Consistent with surveys of Twitter users \cite{Statista}, 63.9\% of the tweets with gender in our analysis were posted by men. 

Our supervised emotion detection approach is based on a RoBERTa-base language model additionally trained on a large sample of English tweets \cite{Barbieri2020}. We included an additional layer in the deep neural network and trained it against all emotion classes of the SemEval '18 emotion dataset \cite{Mohammad2018}, which is used as a benchmark in sentiment analysis competitions that include emotion detection from tweets. In contrast with benchmarks that integrate many more machine learning tasks but only a few emotion classes \cite{Barbieri2020}, we include all emotions in the SemEval dataset to compute scores for the three emotions of interest for our study: sadness, fear, and joy. The classifier achieves high accuracy with Area Under the Curve values between 0.89 and 0.94 (see SI). 

We calculate Twitter signals as daily fractions of tweets that contain an expression of an emotion over all the tweets posted that day. Weekly values are calculated as seven-day rolling averages, using the corresponding window to each week of the YouGov survey. We repeat the analysis for each gender by calculating the same signals but only among the tweets posted by a user of the corresponding gender, excluding tweets from users for whom gender could not be detected. Gender-rescaled signals are calculated as the mean of the signal across genders, which corrects the oversampling of tweets posted by male users by rescaling to a situation in which each gender has the same weight. This approach is based on previous research rescaling social media data \cite{Barbera2016} and works using voter registries \cite{Grinberg2019}.

Following our pre-registration (\url{https://aspredicted.org/blind.php?x=r89nv2}), we analyzed the text of tweets using the anxiety, sadness, and positive affect dictionaries of LIWC  through Brandwatch slightly adapted as in our previous work \cite{Metzler2021}. This way, we obtain a daily count of tweets within the sample that contain at least one emotional term, which we then average over weekly windows aligned with the YouGov survey. Second, we selected a random subsample of $10000$ tweets per day, retrieved their text from Twitter to have an updated dataset, and run our RoBERTa emotion classifier on their text. This gives us an estimate of the prevalence of sadness, fear, and joy for each day as the mean of each emotion score over the tweets of that day.

We use the data provided by YouGov for their weekly UK mood survey \cite{Yougov1}, which started in late June 2019, and the weekly UK satisfaction with life survey \cite{Yougov2}, which started in April 2020. Until July 1st, 2021, these surveys included approximately 2000 responses per week by questioning a panel representative in terms of demographic attributes and political positions \cite{Yougov3}.

For each pair of YouGov emotions and the corresponding text signal on Twitter, we calculate cross-correlation coefficients between the weekly values of each variable. We perform a permutation test with 10.000 permutations of the Twitter time series, while keeping the YouGov time series intact, to provide a null model for comparison with our correlation coefficients. To account for the autocorrelation of the YouGov emotion signals, we perform Detrended Cross Correlation Analysis \cite{Podobnik2008} with a linear trend and a window of 12 weeks. Furthermore, we fit regression models of the value of the YouGov emotion of interest as a function of the corresponding Twitter signal and the previous value of the YouGov emotion. We analyze the results of these regressions by correcting for autocorrelation and heteroscedasticity and test the stationarity of their residuals. The results of these robustness tests are included in the SI.


\begin{thebibliography}{10}
\urlstyle{rm}
\expandafter\ifx\csname url\endcsname\relax
  \def\url#1{\texttt{#1}}\fi
\expandafter\ifx\csname urlprefix\endcsname\relax\def\urlprefix{URL }\fi
\expandafter\ifx\csname doiprefix\endcsname\relax\def\doiprefix{DOI: }\fi
\providecommand{\bibinfo}[2]{#2}
\providecommand{\eprint}[2][]{\url{#2}}

\bibitem{Salganik2019}
\bibinfo{author}{Salganik, M.~J.}
\newblock \emph{\bibinfo{title}{Bit by bit: Social research in the digital
  age}} (\bibinfo{publisher}{Princeton University Press},
  \bibinfo{year}{2019}).

\bibitem{Mohammad2018}
\bibinfo{author}{Mohammad, S.}, \bibinfo{author}{Bravo-Marquez, F.},
  \bibinfo{author}{Salameh, M.} \& \bibinfo{author}{Kiritchenko, S.}
\newblock \bibinfo{title}{Semeval-2018 task 1: Affect in tweets}.
\newblock In \emph{\bibinfo{booktitle}{Proceedings of the 12th international
  workshop on semantic evaluation}}, \bibinfo{pages}{1--17}
  (\bibinfo{year}{2018}).

\bibitem{Goldenberg2020}
\bibinfo{author}{Goldenberg, A.}, \bibinfo{author}{Garcia, D.},
  \bibinfo{author}{Halperin, E.} \& \bibinfo{author}{Gross, J.~J.}
\newblock \bibinfo{journal}{\bibinfo{title}{Collective emotions}}.
\newblock {\emph{\JournalTitle{Current Directions in Psychological Science}}}
  \textbf{\bibinfo{volume}{29}}, \bibinfo{pages}{154--160}
  (\bibinfo{year}{2020}).

\bibitem{Golder2011}
\bibinfo{author}{Golder, S.~A.} \& \bibinfo{author}{Macy, M.~W.}
\newblock \bibinfo{journal}{\bibinfo{title}{Diurnal and seasonal mood vary with
  work, sleep, and daylength across diverse cultures}}.
\newblock {\emph{\JournalTitle{Science}}} \textbf{\bibinfo{volume}{333}},
  \bibinfo{pages}{1878--1881} (\bibinfo{year}{2011}).

\bibitem{Dodds2011}
\bibinfo{author}{Dodds, P.~S.}, \bibinfo{author}{Harris, K.~D.},
  \bibinfo{author}{Kloumann, I.~M.}, \bibinfo{author}{Bliss, C.~A.} \&
  \bibinfo{author}{Danforth, C.~M.}
\newblock \bibinfo{journal}{\bibinfo{title}{Temporal patterns of happiness and
  information in a global social network: Hedonometrics and twitter}}.
\newblock {\emph{\JournalTitle{PloS one}}} \textbf{\bibinfo{volume}{6}},
  \bibinfo{pages}{e26752} (\bibinfo{year}{2011}).

\bibitem{Beasley2016}
\bibinfo{author}{Beasley, A.}, \bibinfo{author}{Mason, W.} \&
  \bibinfo{author}{Smith, E.}
\newblock \bibinfo{journal}{\bibinfo{title}{Inferring emotions and
  self-relevant domains in social media: Challenges and future directions.}}
\newblock {\emph{\JournalTitle{Translational Issues in Psychological Science}}}
  \textbf{\bibinfo{volume}{2}}, \bibinfo{pages}{238} (\bibinfo{year}{2016}).

\bibitem{Elayan2020}
\bibinfo{author}{Elayan, S.} \emph{et~al.}
\newblock \bibinfo{title}{The stresscapes ontology system: Detecting and
  measuring stress on social media}.
\newblock In \emph{\bibinfo{booktitle}{7th European Conference on Social Media
  ECSM 2020}}, \bibinfo{pages}{74} (\bibinfo{year}{2020}).

\bibitem{Jaidka2020}
\bibinfo{author}{Jaidka, K.} \emph{et~al.}
\newblock \bibinfo{journal}{\bibinfo{title}{Estimating geographic subjective
  well-being from twitter: A comparison of dictionary and data-driven language
  methods}}.
\newblock {\emph{\JournalTitle{Proceedings of the National Academy of
  Sciences}}} \textbf{\bibinfo{volume}{117}}, \bibinfo{pages}{10165--10171}
  (\bibinfo{year}{2020}).

\bibitem{Galesic2021}
\bibinfo{author}{Galesic, M.} \emph{et~al.}
\newblock \bibinfo{journal}{\bibinfo{title}{Human social sensing is an untapped
  resource for computational social science}}.
\newblock {\emph{\JournalTitle{Nature}}} \bibinfo{pages}{1--9}
  (\bibinfo{year}{2021}).

\bibitem{Lazer2021}
\bibinfo{author}{Lazer, D.} \emph{et~al.}
\newblock \bibinfo{journal}{\bibinfo{title}{Meaningful measures of human
  society in the twenty-first century}}.
\newblock {\emph{\JournalTitle{Nature}}} \bibinfo{pages}{1--8}
  (\bibinfo{year}{2021}).

\bibitem{Wagner2021}
\bibinfo{author}{Wagner, C.} \emph{et~al.}
\newblock \bibinfo{journal}{\bibinfo{title}{Measuring algorithmically infused
  societies}}.
\newblock {\emph{\JournalTitle{Nature}}} \bibinfo{pages}{1--6}
  (\bibinfo{year}{2021}).

\bibitem{Dukes2021rise}
\bibinfo{author}{Dukes, D.} \emph{et~al.}
\newblock \bibinfo{journal}{\bibinfo{title}{The rise of affectivism}}.
\newblock {\emph{\JournalTitle{Nature Human Behaviour}}} \bibinfo{pages}{1--5}
  (\bibinfo{year}{2021}).

\bibitem{Iyengar2019}
\bibinfo{author}{Iyengar, S.}, \bibinfo{author}{Lelkes, Y.},
  \bibinfo{author}{Levendusky, M.}, \bibinfo{author}{Malhotra, N.} \&
  \bibinfo{author}{Westwood, S.~J.}
\newblock \bibinfo{journal}{\bibinfo{title}{The origins and consequences of
  affective polarization in the united states}}.
\newblock {\emph{\JournalTitle{Annual Review of Political Science}}}
  \textbf{\bibinfo{volume}{22}}, \bibinfo{pages}{129--146}
  (\bibinfo{year}{2019}).

\bibitem{Finkel2020}
\bibinfo{author}{Finkel, E.~J.} \emph{et~al.}
\newblock \bibinfo{journal}{\bibinfo{title}{Political sectarianism in
  america}}.
\newblock {\emph{\JournalTitle{Science}}} \textbf{\bibinfo{volume}{370}},
  \bibinfo{pages}{533--536} (\bibinfo{year}{2020}).

\bibitem{Martel2020}
\bibinfo{author}{Martel, C.}, \bibinfo{author}{Pennycook, G.} \&
  \bibinfo{author}{Rand, D.~G.}
\newblock \bibinfo{journal}{\bibinfo{title}{Reliance on emotion promotes belief
  in fake news}}.
\newblock {\emph{\JournalTitle{Cognitive research: principles and
  implications}}} \textbf{\bibinfo{volume}{5}}, \bibinfo{pages}{1--20}
  (\bibinfo{year}{2020}).

\bibitem{Rathje2021}
\bibinfo{author}{Rathje, S.}, \bibinfo{author}{Van~Bavel, J.~J.} \&
  \bibinfo{author}{van~der Linden, S.}
\newblock \bibinfo{journal}{\bibinfo{title}{Out-group animosity drives
  engagement on social media}}.
\newblock {\emph{\JournalTitle{Proceedings of the National Academy of
  Sciences}}} \textbf{\bibinfo{volume}{118}} (\bibinfo{year}{2021}).

\bibitem{Chancellor2020}
\bibinfo{author}{Chancellor, S.} \& \bibinfo{author}{De~Choudhury, M.}
\newblock \bibinfo{journal}{\bibinfo{title}{Methods in predictive techniques
  for mental health status on social media: a critical review}}.
\newblock {\emph{\JournalTitle{NPJ digital medicine}}}
  \textbf{\bibinfo{volume}{3}}, \bibinfo{pages}{1--11} (\bibinfo{year}{2020}).

\bibitem{Garcia2019}
\bibinfo{author}{Garcia, D.} \& \bibinfo{author}{Rim{\'e}, B.}
\newblock \bibinfo{journal}{\bibinfo{title}{Collective emotions and social
  resilience in the digital traces after a terrorist attack}}.
\newblock {\emph{\JournalTitle{Psychological science}}}
  \textbf{\bibinfo{volume}{30}}, \bibinfo{pages}{617--628}
  (\bibinfo{year}{2019}).

\bibitem{Fan2019}
\bibinfo{author}{Fan, R.} \emph{et~al.}
\newblock \bibinfo{journal}{\bibinfo{title}{The minute-scale dynamics of online
  emotions reveal the effects of affect labeling}}.
\newblock {\emph{\JournalTitle{Nature human behaviour}}}
  \textbf{\bibinfo{volume}{3}}, \bibinfo{pages}{92--100}
  (\bibinfo{year}{2019}).

\bibitem{Yarrington2021}
\bibinfo{author}{Yarrington, J.~S.} \emph{et~al.}
\newblock \bibinfo{journal}{\bibinfo{title}{Impact of the covid-19 pandemic on
  mental health among 157,213 americans}}.
\newblock {\emph{\JournalTitle{Journal of Affective Disorders}}}
  \textbf{\bibinfo{volume}{286}}, \bibinfo{pages}{64--70}
  (\bibinfo{year}{2021}).

\bibitem{Gayo2013}
\bibinfo{author}{Gayo-Avello, D.}
\newblock \bibinfo{journal}{\bibinfo{title}{A meta-analysis of state-of-the-art
  electoral prediction from twitter data}}.
\newblock {\emph{\JournalTitle{Social Science Computer Review}}}
  \textbf{\bibinfo{volume}{31}}, \bibinfo{pages}{649--679}
  (\bibinfo{year}{2013}).

\bibitem{Pennebaker2015}
\bibinfo{author}{Pennebaker, J.~W.}, \bibinfo{author}{Boyd, R.~L.},
  \bibinfo{author}{Jordan, K.} \& \bibinfo{author}{Blackburn, K.}
\newblock \bibinfo{title}{The development and psychometric properties of
  liwc2015}.
\newblock \bibinfo{type}{Tech. Rep.} (\bibinfo{year}{2015}).

\bibitem{Nilizadeh2016}
\bibinfo{author}{Nilizadeh, S.} \emph{et~al.}
\newblock \bibinfo{title}{Twitter's glass ceiling: The effect of perceived
  gender on online visibility}.
\newblock In \emph{\bibinfo{booktitle}{Proceedings of the International AAAI
  Conference on Web and Social Media}}, vol.~\bibinfo{volume}{10}
  (\bibinfo{year}{2016}).

\bibitem{chaplin_gender_2013}
\bibinfo{author}{Chaplin, T.~M.} \& \bibinfo{author}{Aldao, A.}
\newblock \bibinfo{journal}{\bibinfo{title}{Gender differences in emotion
  expression in children: {A} meta-analytic review}}.
\newblock {\emph{\JournalTitle{Psychological Bulletin}}}
  \textbf{\bibinfo{volume}{139}}, \bibinfo{pages}{735--765},
  \doiprefix\url{10.1037/a0030737} (\bibinfo{year}{2013}).
\newblock \bibinfo{note}{Place: US Publisher: American Psychological
  Association}.

\bibitem{mclean_brave_2009}
\bibinfo{author}{McLean, C.~P.} \& \bibinfo{author}{Anderson, E.~R.}
\newblock \bibinfo{journal}{\bibinfo{title}{Brave men and timid women? {A}
  review of the gender differences in fear and anxiety}}.
\newblock {\emph{\JournalTitle{Clinical Psychology Review}}}
  \textbf{\bibinfo{volume}{29}}, \bibinfo{pages}{496--505},
  \doiprefix\url{10.1016/j.cpr.2009.05.003} (\bibinfo{year}{2009}).
\newblock \bibinfo{note}{Place: Oxford Publisher: Pergamon-Elsevier Science Ltd
  WOS:000269173100004}.

\bibitem{whittle_sex_2011}
\bibinfo{author}{Whittle, S.}, \bibinfo{author}{Yücel, M.},
  \bibinfo{author}{Yap, M. B.~H.} \& \bibinfo{author}{Allen, N.~B.}
\newblock \bibinfo{journal}{\bibinfo{title}{Sex differences in the neural
  correlates of emotion: {Evidence} from neuroimaging}}.
\newblock {\emph{\JournalTitle{Biological Psychology}}}
  \textbf{\bibinfo{volume}{87}}, \bibinfo{pages}{319--333},
  \doiprefix\url{10.1016/j.biopsycho.2011.05.003} (\bibinfo{year}{2011}).

\bibitem{Metzler2021}
\bibinfo{author}{Metzler, H.} \emph{et~al.}
\newblock \bibinfo{journal}{\bibinfo{title}{Collective emotions during the
  covid-19 outbreak}}.
\newblock  (\bibinfo{year}{2021}).

\bibitem{Mellon2017}
\bibinfo{author}{Mellon, J.} \& \bibinfo{author}{Prosser, C.}
\newblock \bibinfo{journal}{\bibinfo{title}{Twitter and facebook are not
  representative of the general population: Political attitudes and
  demographics of british social media users}}.
\newblock {\emph{\JournalTitle{Research \& Politics}}}
  \textbf{\bibinfo{volume}{4}}, \bibinfo{pages}{2053168017720008}
  (\bibinfo{year}{2017}).

\bibitem{Conrad2019}
\bibinfo{author}{Conrad, F.~G.} \emph{et~al.}
\newblock \bibinfo{journal}{\bibinfo{title}{Social media as an alternative to
  surveys of opinions about the economy}}.
\newblock {\emph{\JournalTitle{Social Science Computer Review}}}
  \bibinfo{pages}{0894439319875692} (\bibinfo{year}{2019}).

\bibitem{Munger2019}
\bibinfo{author}{Munger, K.}
\newblock \bibinfo{journal}{\bibinfo{title}{The limited value of non-replicable
  field experiments in contexts with low temporal validity}}.
\newblock {\emph{\JournalTitle{Social Media+ Society}}}
  \textbf{\bibinfo{volume}{5}}, \bibinfo{pages}{2056305119859294}
  (\bibinfo{year}{2019}).

\bibitem{Brandwatch2020}
\bibinfo{author}{{Crimson Hexagon}}.
\newblock \bibinfo{title}{Location methodology} (\bibinfo{year}{2020}).

\bibitem{Brandwatch2020b}
\bibinfo{author}{{Crimson Hexagon}}.
\newblock \bibinfo{title}{Forsight: User guide} (\bibinfo{year}{2020}).

\bibitem{Statista}
\bibinfo{author}{Statista}.
\newblock \bibinfo{title}{Distribution of twitter users worldwide as of april
  2021, by gender} (\bibinfo{year}{2021}).

\bibitem{Barbieri2020}
\bibinfo{author}{Barbieri, F.}, \bibinfo{author}{Camacho-Collados, J.},
  \bibinfo{author}{Anke, L.~E.} \& \bibinfo{author}{Neves, L.}
\newblock \bibinfo{title}{Tweeteval: Unified benchmark and comparative
  evaluation for tweet classification}.
\newblock In \emph{\bibinfo{booktitle}{Proceedings of the 2020 Conference on
  Empirical Methods in Natural Language Processing: Findings}},
  \bibinfo{pages}{1644--1650} (\bibinfo{year}{2020}).

\bibitem{Barbera2016}
\bibinfo{author}{Barber{\'a}, P.}
\newblock \bibinfo{journal}{\bibinfo{title}{Less is more? how demographic
  sample weights can improve public opinion estimates based on twitter data}}.
\newblock {\emph{\JournalTitle{Work Pap NYU}}}  (\bibinfo{year}{2016}).

\bibitem{Grinberg2019}
\bibinfo{author}{Grinberg, N.}, \bibinfo{author}{Joseph, K.},
  \bibinfo{author}{Friedland, L.}, \bibinfo{author}{Swire-Thompson, B.} \&
  \bibinfo{author}{Lazer, D.}
\newblock \bibinfo{journal}{\bibinfo{title}{Fake news on twitter during the
  2016 us presidential election}}.
\newblock {\emph{\JournalTitle{Science}}} \textbf{\bibinfo{volume}{363}},
  \bibinfo{pages}{374--378} (\bibinfo{year}{2019}).

\bibitem{Yougov1}
\bibinfo{author}{{YouGov}}.
\newblock \bibinfo{title}{Britain's mood, measured weekly}
  (\bibinfo{year}{2021}).

\bibitem{Yougov2}
\bibinfo{author}{{YouGov}}.
\newblock \bibinfo{title}{Life satisfaction, measured weekly}
  (\bibinfo{year}{2021}).

\bibitem{Yougov3}
\bibinfo{author}{{YouGov}}.
\newblock \bibinfo{title}{Panel methodology} (\bibinfo{year}{2021}).

\bibitem{Podobnik2008}
\bibinfo{author}{Podobnik, B.} \& \bibinfo{author}{Stanley, H.~E.}
\newblock \bibinfo{journal}{\bibinfo{title}{Detrended cross-correlation
  analysis: a new method for analyzing two nonstationary time series}}.
\newblock {\emph{\JournalTitle{Physical review letters}}}
  \textbf{\bibinfo{volume}{100}}, \bibinfo{pages}{084102}
  (\bibinfo{year}{2008}).

\end{thebibliography}

\begin{thebibliography}{1}
\urlstyle{rm}
\expandafter\ifx\csname url\endcsname\relax
  \def\url#1{\texttt{#1}}\fi
\expandafter\ifx\csname urlprefix\endcsname\relax\def\urlprefix{URL }\fi
\expandafter\ifx\csname doiprefix\endcsname\relax\def\doiprefix{DOI: }\fi
\providecommand{\bibinfo}[2]{#2}
\providecommand{\eprint}[2][]{\url{#2}}

\bibitem{Barbieri2020}
\bibinfo{author}{Barbieri, F.}, \bibinfo{author}{Camacho-Collados, J.},
  \bibinfo{author}{Anke, L.~E.} \& \bibinfo{author}{Neves, L.}
\newblock \bibinfo{title}{Tweeteval: Unified benchmark and comparative
  evaluation for tweet classification}.
\newblock In \emph{\bibinfo{booktitle}{Proceedings of the 2020 Conference on
  Empirical Methods in Natural Language Processing: Findings}},
  \bibinfo{pages}{1644--1650} (\bibinfo{year}{2020}).

\bibitem{Mohammad2018}
\bibinfo{author}{Mohammad, S.}, \bibinfo{author}{Bravo-Marquez, F.},
  \bibinfo{author}{Salameh, M.} \& \bibinfo{author}{Kiritchenko, S.}
\newblock \bibinfo{title}{Semeval-2018 task 1: Affect in tweets}.
\newblock In \emph{\bibinfo{booktitle}{Proceedings of the 12th international
  workshop on semantic evaluation}}, \bibinfo{pages}{1--17}
  (\bibinfo{year}{2018}).

\bibitem{Andrews1992}
\bibinfo{author}{Andrews, D.~W.} \& \bibinfo{author}{Monahan, J.~C.}
\newblock \bibinfo{journal}{\bibinfo{title}{An improved heteroskedasticity and
  autocorrelation consistent covariance matrix estimator}}.
\newblock {\emph{\JournalTitle{Econometrica: Journal of the Econometric
  Society}}} \bibinfo{pages}{953--966} (\bibinfo{year}{1992}).

\bibitem{Kwiatkowski1992}
\bibinfo{author}{Kwiatkowski, D.}, \bibinfo{author}{Phillips, P.~C.},
  \bibinfo{author}{Schmidt, P.} \& \bibinfo{author}{Shin, Y.}
\newblock \bibinfo{journal}{\bibinfo{title}{Testing the null hypothesis of
  stationarity against the alternative of a unit root: How sure are we that
  economic time series have a unit root?}}
\newblock {\emph{\JournalTitle{Journal of econometrics}}}
  \textbf{\bibinfo{volume}{54}}, \bibinfo{pages}{159--178}
  (\bibinfo{year}{1992}).

\bibitem{derczynski2017}
\bibinfo{author}{Derczynski, L.}, \bibinfo{author}{Nichols, E.},
  \bibinfo{author}{van Erp, M.} \& \bibinfo{author}{Limsopatham, N.}
\newblock \bibinfo{title}{Results of the {WNUT}2017 shared task on novel and
  emerging entity recognition}.
\newblock In \emph{\bibinfo{booktitle}{Proceedings of the 3rd Workshop on Noisy
  User-generated Text}}, \bibinfo{pages}{140--147},
  \doiprefix\url{10.18653/v1/W17-4418} (\bibinfo{publisher}{Association for
  Computational Linguistics}, \bibinfo{address}{Copenhagen, Denmark},
  \bibinfo{year}{2017}).

\end{thebibliography}

\section*{Acknowledgements}
The research leading to these results received funding from the Vienna Science and Technology Fund through the project "Emotional Well-Being in the Digital Society" (Grant No. VRG16-005). We would like to thank Jan-Emmanuel De Neve for useful comments.

\section*{Author contributions statement}
D.G., J.L. and H.M. conceived the research, D.G. collected data, M.P. developed text analysis methods and performed data analysis, D.G. performed statistical analysis and wrote the manuscript. All authors reviewed the manuscript. 

\section*{Additional information}
Accompanying data can be found in the following github repository: \url{https://github.com/dgarcia-eu/MacroEmotions}. The repository will be updated in the future with codes to reproduce the final results of the article and with the emotion classification model.

\newpage 

\section{Supplementary Information}

\subsection{Supervised emotion detection in tweet text}

We designed a supervised emotion classifier from tweet text based on the RoBERTa-base language model that was additionally retrained with a dataset of 60 Million English tweets retrieved through Twitter's streaming API between May 2018 and August 2019 \cite{Barbieri2020}. We extended the RoBERTa neural network by adding a last layer to perform a classification of tweets into distinct emotion labels. We used the original dataset of the SemEval '18 emotion classification competition \cite{Mohammad2018}, rehydrating tweets in 2020 to have a sample of XXX tweets for training and XXX tweets for testing. We chose to fit our own model against the original SemEval dataset rather than using the TweetEval benchmark because the TweetEval benchmark only includes a short list of frequent emotion labels that does not include fear, which is one of the emotions we include in our study.

\begin{figure}[h]
\centering
\includegraphics[width=\linewidth]{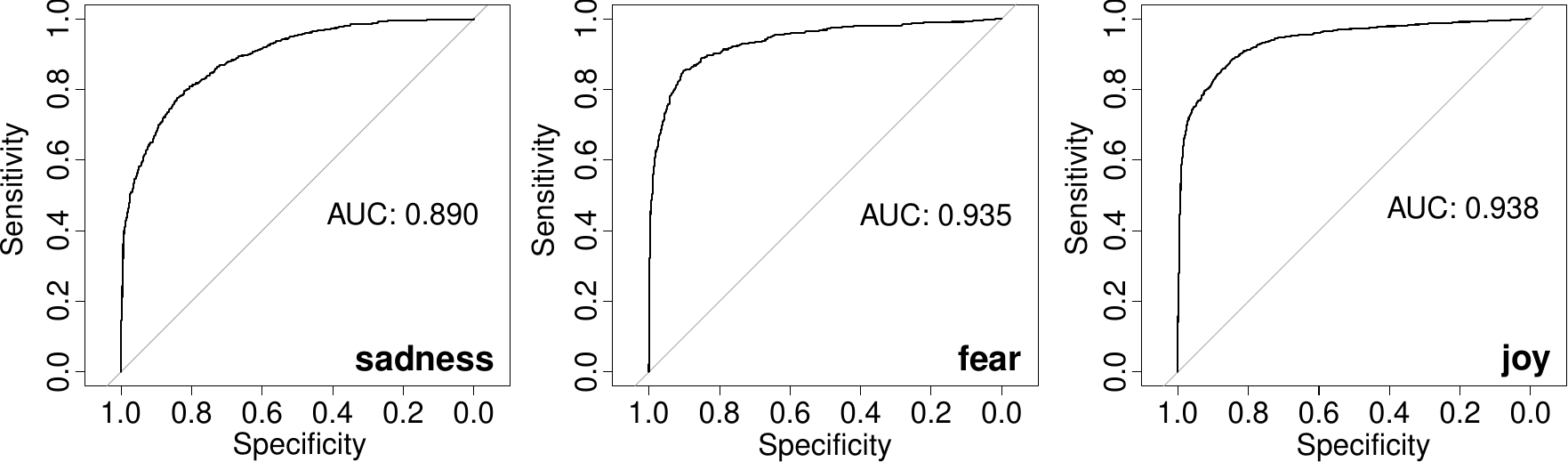}
\caption{Receiver Operator Characteristic (ROC) curves of emotion detection in the test dataset of TweetEval with reported Area Under the Curve (AUC).}
\label{fig:AUC}
\end{figure}

The classifier returns a score for each emotion between zero and one. The Receiver Operator Characteristic curves of the three classes over the test dataset is shown on Figure \ref{fig:AUC}. The classifier achieves an Area Under the Curve (AUC) around 0.9 for all three classes, showing good predictive power with respect to detecting emotion labels in tweets. 

\subsection{Pre-registration and additional analyses}

On October 30th, 2020, we pre-registered the central hypotheses and methods of our study: \url{https://aspredicted.org/blind.php?x=r89nv2}. We had performed a cursory analysis of survey and Twitter data up to August 2020, and thus registered our analysis in two parts: a historical part with data until October 31st, 2020, and a predictive part with data from November 1st, 2020. On the date of the pre-registration, we had not collected yet the data after August 2020, leaving the final data collection for a date after the prediction period in 2021. We initially planned our data collection to happen in March 2021, but personal reasons of the lead author independent of the study (he became a father for the first time) delayed the data collection until early July 2021. As a result, we count with eight months in the prediction period as opposed to the previously planned five, but since we had no power analysis to determine the sample size a priori, we decided to use all data available up to the study date to provide completeness to the analysis.

We only pre-registered the hypotheses with respect to dictionary methods, as we implemented the supervised classifier in late 2020. Thus, the supervised part of the study can be considered a reanalysis of the same data with the same hypothesis but with an alternative method that shows high reliability in the classification task. With respect to gender, we structured the pre-registration as including a gender-agnostic analysis and then a gender-based analysis for stratified samples for male and female users and survey respondents. The gender-rescaled version of our analysis is thus a way to summarize the gender-informed results of our pre-registration, with the precise results for each gender reported on Table \ref{tab:3} of this Supplementary Information.

The pre-registration contained an exploratory component to test the correlation between the frequency of tweets with explicit expressions of emotion and the 12 emotional states of the YouGov survey. We performed that analysis with the same statistical methods as for our hypotheses. The analysis of the Satisfaction With Life survey of YouGov and of third-person references were not part of our pre-registration. Those additional analyses were included later to understand better the limitations of this method and the reasons why social media emotions correlate so consistently with survey responses.

\newpage
\subsection{Details on correlations and model fits}

\begin{table}[ht]
\centering
\begin{tabular}{|c|c|c|c|c|c|}
\hline
Emotion (survey) & Twitter signal & $r_1$ (historical) &  $r_2$ (predicted)    \\
\hline
Sad & LIWC sad & $0.714^{***} \quad [0.577, 0.812]$ & $0.633^{***} \quad [0.379, 0.798]$\\
Sad & RoBERTa sad & $0.649^{***} \quad [0.49, 0.766]$ & $0.64^{***} \quad [0.39, 0.802]$\\
Scared & LIWC anxiety & $0.802^{***} \quad [0.7, 0.872]$ & $0.257 (n.s.) \quad [-0.084, 0.543]$\\
Scared & RoBERTa fear & $0.798^{***} \quad [0.694, 0.87]$ & $0.017 (n.s.) \quad [-0.318, 0.348]$\\
Happy & LIWC positive & $0.297^{*} \quad [0.068, 0.496]$ & $-0.093 (n.s.) \quad [-0.413, 0.248]$\\
Happy & RoBERTa joy & $0.67^{***} \quad [0.517, 0.781]$ & $0.531^{**} \quad [0.24, 0.734]$\\
\hline
\end{tabular}
\caption{\label{tab:1} Correlation coefficients of the weekly percentage of emotions in YouGov and a LIWC or RoBERTa estimator based on Twitter data without gender rescaling. 95\% Confidence Intervals are reported. $^{\cdot} p<0.1, ^{*} p<0.05, ^{**} p<0.01, ^{***} p<0.001$}
\end{table}

We perform HAC tests \cite{Andrews1992} to provide a robust assessment of the informativeness of Twitter signals when autocorrelation and heteroscedasticity might be present. Finally, we check that the residuals of each of these models can be considered stationary with KPSS tests \cite{Kwiatkowski1992}.

\begin{table}[ht]
\centering
\begin{tabular}{|c|c|c|c|c|c|c|}
\hline
Emotion (survey) & Twitter signal & DCCA & $\beta$ & KPSS  \\
\hline
Sad & LIWC sad & $0.64^{***}$  & $0.602^{***}$ & $p>0.1$  \\
Sad & RoBERTa sad & $0.611^{***}$  & $0.515^{***}$ & $p>0.1$ \\
Scared & LIWC anxiety & $0.788^{***}$  & $0.472^{***}$ & $p>0.05$ \\
Scared & RoBERTa fear & $0.798^{***}$  & $0.436^{*}$ & $p>0.1$   \\
Happy & LIWC positive & $0.271 (n.s)$  & $0.175(n.s)$ & $p>0.1$  \\
Happy & RoBERTa joy & $0.656^{***}$  & $0.313^{**}$ & $p>0.1$ \\
\hline
\end{tabular}
\caption{\label{tab:2} Robustness analysis results for gender-rescaled Twitter signals. Detrended Cross-Correlation Analysis (DCCA) p-values are based on $10000$ permutations. $\beta$ coefficients of time series model with additional lags have p-values corrected for heteroscedasticity and autocorrelation.  Kwiatkowski-Phillips-Schmidt-Shin (KPSS) tests of residuals report p values above 0.05 (i.e. cannot reject stationarity of residuals). $^{\cdot} p<0.1, ^{*} p<0.05, ^{**} p<0.01, ^{***} p<0.001$}
\end{table}

\begin{table}[ht]
\centering
\begin{tabular}{|c|c|c|c|c|c|c|c|c|c|}
\hline
Gender & Emotion & LIWC signal & $r_1$ (historical) &  $r_2$ (predicted)  & DCCA & $\beta$ & KPSS  \\
\hline
Male & Sad & Sadness & $0.532^{***} [0.342, 0.681]$ & $ 0.475^{***} [0.168, 0.698]$ & $0.609^{***}$  & $ 0.482^{***}$ & $p>0.1$  \\
Female & Sad & Sadness & $0.68^{***} [0.531, 0.788]$ & $0.741^{***} [0.542, 0.862]$ & $0.5^{***}$  & $ 0.527^{***}$ & $p>0.1$  \\
\hline
Male & Scared & Anxiety & $0.684^{***} [0.536, 0.791]$ & $ 0.478^{***} [0.172, 0.7]$ & $0.724^{***}$  & $ 0.454^{***}$ & $p>0.1$  \\
Female & Scared & Anxiety & $0.81^{***} [0.711, 0.878]$ & $0.365^{***} [ 0.036, 0.623]$ & $ 0.811^{***}$  & $0.479^{***}$ & $p>0.1$  \\
\hline
Male & Happy & Positive Affect & $0.303^{*} [0.075, 0.501]$ & $0.015 (n.s.) [-0.32, 0.347]$ & $0.379^{*}$  & $0.196^{\cdot}$ & $p>0.1$  \\
Female & Happy & Positive Affect & $ 0.239^{*} [0.006, 0.447]$ & $ 0.019 (n.s.) [-0.316, 0.35]$ & $0.127 (n.s.)$  & $0.156^{\cdot}$ & $p>0.1$  \\
\hline
\end{tabular}
\caption{\label{tab:3} Correlation coefficients of the weekly percentage of emotions in YouGov and a LIWC or RoBERTa estimator based on Twitter data stratified by gender. 95\% Confidence Intervals are reported. Robustness analysis for each stratum and signal are based on the same metrics as in Table \label{tab:2}. $^{\cdot} p<0.1, ^{*} p<0.05, ^{**} p<0.01, ^{***} p<0.001$}
\end{table}

\newpage

\subsection{Details on results with 12 emotions}
\begin{figure}[h]
\centering
\includegraphics[width=0.5\linewidth]{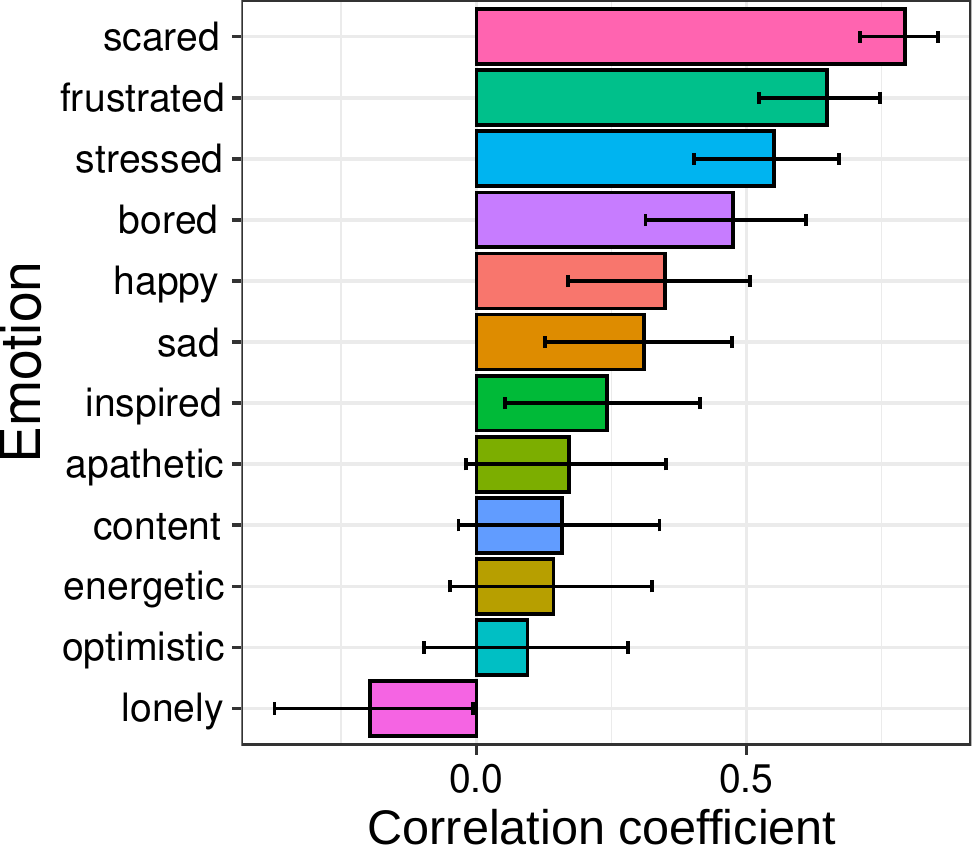}
\caption{Correlation coefficients between the percent of weekly responses for an emotion in the YouGov survey and the weekly volume of tweets expliclitely reporting that emotion without applying gender rescaling. Error bars show 95\% confidence intervals.}
\label{fig:2}
\end{figure}

\begin{table}[ht]
\centering
\begin{tabular}{|c|c|c|c|c|}
\hline
Emotion & Correlation coefficient & Correlation coefficient (gender-rescaled) & Mean \% in YouGov & Mean num. tweets \\
\hline
happy & 0.35 [0.17, 0.507] & 0.367 [0.19, 0.522] & 44\% & 7560 \\
sad & 0.31 [0.127, 0.473]  & 0.242 [0.054, 0.414] & 26\% & 2196 \\
scared & 0.794 [0.711, 0.855] & 0.752 [0.655, 0.824] & 12\% & 1195 \\
bored & 0.475 [0.313, 0.611]  & 0.488 [0.327, 0.621] & 26\% & 973 \\
stressed & 0.552 [0.403, 0.672] & 0.563 [ 0.416, 0.680] & 40\% & 340 \\
optimistic & 0.095 [-0.098, 0.281] & 0.071 [-0.121, 0.258]  & 21\% & 215 \\
inspired & 0.242 [0.053, 0.413]  & 0.236 [0.047, 0.408] & 10\% & 202 \\
frustrated & 0.649 [0.523, 0.747] & 0.612 [0.476, 0.719] & 38\% & 194 \\
lonely & -0.197 [-0.374, -0.007] & -0.123 [-0.307, 0.069]   & 18\% & 183 \\
content & 0.158 [-0.033, 0.339] & 0.094 [-0.099, 0.279] & 26\% & 62 \\
energetic & 0.143 [-0.049, 0.325] & 0.189 [-0.001, 0.367] & 14\% & 11 \\
apathetic & 0.172 [-0.019, 0.351] & 0.178 [-0.013, 0.356] & 19\% & 7 \\
\hline
\end{tabular}
\caption{\label{tab:yg} Correlation coefficients of the weekly percentage of emotions in YouGov and a text estimator based on explicit first-person statements about the emotion. 95\% Confidence Intervals are reported. The last two columns report the mean of the weekly percentage of each emotion in YouGov and the mean of the weekly number of tweets that contain an explicit mention to that emotion. Emotions with substantial weekly numbers of tweets display positive correlations with YouGov percentages.}
\end{table}

\newpage

\subsection{Third-person references in emotion tweets}

We detected third-person references in tweets by matching whether a tweet contains one of the words in this list: "they", "them", "their", "he", "him" "his", "she", "her", "hers". We calculated the daily fraction of tweets including at least one of these words from among all tweets in our analysis as a baseline calculation, to then repeat the calculation but only among the tweets that contain a LIWC term from the tree dictionaries. The frequencies and comparisons between the baseline and each emotion case are shown on Table \ref{tab:3rd}. All comparisons between frequencies are statistically significant in  $\chi^2$ tests of difference in proportions.

\begin{table}[ht]
\centering
\begin{tabular}{|c|c|c|c|c|}
\hline
Sample & fraction &  Sample & fraction & \% difference \\ \hline
All tweets (baseline) &  17.3\% & - & - & -  \\
With anxiety terms &  29.3\% & Without anxiety terms & 16.7\% & 74.85\%\\
With sad terms &  27\% & Without sad terms & 16.66\% &  62.12\%\\
With positive terms &  20.3\% & Without positive terms & 15\% & 34.97\%\\
\hline
\end{tabular}
\caption{\label{tab:3rd} Proportions of tweets with third-person pronouns from among all tweets in the dataset and from among tweets including at least on term of each considered LIWC emotion class and without any term of that class. $\chi^2$ tests of all proportion comparisons have p$<0.0001$}
\end{table}

\begin{table}[ht]
\centering
\begin{tabular}{|c|c|c|c|c|}
\hline
Sample & fraction &  Sample & fraction & \% difference \\ \hline
All tweets (baseline) &  36.42\% & - & - & -  \\
With anxiety terms &  37.51\% & Without anxiety terms & 36.37\% & 3.11\%\\
With sad terms &  41.94\% & Without sad terms & 36.07\% &  16.27\%\\
With positive terms &  39.82\% & Without positive terms & 33.07\% & 20.41\%\\
\hline
\end{tabular}
\caption{\label{tab:3rd} Proportions of tweets with at least one person identified with named entity recognition trained with the WNUT
17 dataset \cite{derczynski2017} from among all tweets in the dataset and from among tweets including at least on term of each considered LIWC emotion class and without any term of that class. $\chi^2$ tests of all proportion comparisons have p$<0.0001$}
\end{table}

\end{document}